\documentclass[journal]{IEEEtran}

\usepackage{blindtext}
\usepackage{graphicx}
\usepackage{blindtext, graphicx}
\usepackage{color,soul}
\usepackage{mwe}
\usepackage{caption}
\usepackage{subcaption}
\usepackage{graphicx}
\usepackage{cite}
\usepackage{balance}
\usepackage{setspace}

\def\NoNumber#1{{\def\alglinenumber##1{}\State #1}\addtocounter{ALG@line}{-1}}
\ifCLASSINFOpdf
\else
\fi
\usepackage{amsmath,amssymb}
\usepackage{amsmath}
\usepackage{algorithm}
\usepackage[noend]{algpseudocode}
\usepackage{algorithm,algpseudocode}
\makeatletter
\newcommand{\algmargin}{\the\ALG@thistlm}
\makeatother
\newlength{\whilewidth}
\algdef{SE}[parWHILE]{parWhile}{EndparWhile}[1]
  {\parbox[t]{\dimexpr\linewidth-\algmargin}{%
     \hangindent\whilewidth\strut\algorithmicwhile\ #1\ \algorithmicdo\strut}}{\algorithmicend\ \algorithmicwhile}%
\algnewcommand{\parState}[1]{\State%
  \parbox[t]{\dimexpr\linewidth-\algmargin}{\strut #1\strut}}

\begin{document}

\title{Joint Power Control, Channel Assignment and Cell Association in Heterogeneous Cellular Networks}

\author{Shahla~Mohsenifard\IEEEmembership{},       Ahmad~R.~Sharafat,~\IEEEmembership{Life Senior Member,~IEEE},~and Halim~Yanikomeroglu~\IEEEmembership{Fellow,~IEEE}
\thanks{S. Mohsenifard and A. R. Sharafat are with the Faculty of ECE, Tarbiat Modares University, Tehran, Iran. H. Yanikomeroglu is with the Department of SCE, Carleton University, Ottawa, Ontario, Canada.

Corresponding author is A. R. Sharafat (email: sharafat@ieee.org).}
}


\maketitle

\begin{abstract}
Heterogeneous network (HetNet) is a promising concept for increasing capacity and alleviating spectrum scarcity. In HetNets, however, channel assignment and transmit power control affect the distribution of users among base stations. We present a novel scheme to maximize the uplink sum rate in two-tier HetNets with one macrocell and several femtocells, where the transmit power of each user is bounded and at least one channel is assigned to each user. We divide the problem into two sub-problems: one for channel assignment and one for transmit power control, and solve them by iteratively alternating between the two. Our scheme is convergent, and yields a transmission rate above 6 bps/Hz for almost 50\% of users as compared to the same for 10\% of users in SINR-based schemes. When users are in cell boundaries, the average transmission rate for fractional frequency reuse is up to 20\% more compared to the conventional full frequency reuse. This is due to reduced transmit power resulting in less interference.
\end{abstract}


\IEEEpeerreviewmaketitle

\vspace{3 mm}
\section{Introduction}\label{Introduction}
\IEEEPARstart{T}{he phenomenal growth} in mobile data calls for innovative approaches for more efficient use of the limited resources. Heterogeneous networks (HetNets) consisting of high-power macro base-stations (BSs) and low-power small BSs, including microcells, relays, picocells and femtocells, is a promising concept for enhancing frequency reuse and reducing transmit power levels. In this paradigm, users receive higher signal levels from macro BSs as compared to small BSs, and SINR levels in the downlink (DL) and uplink (UL) are not the same (asymmetric SINR). As such, BS association based on highest DL SINR, as in macro-cell only networks is not suitable for HetNets. In HetNets, there is a need to balance the load among different cells and reduce intercell interference, which can be achieved by UL transmit power control.  

Recent research on cell association and power control is mostly based on downlink (DL) metrics \cite{Quser,Qon,Singh}, which may not be optimal for UL  due to asymmetry of metrics in DL and UL \cite{Jan}. In \cite{Jan}, the long-term average data rate is maximized and the quality of service (QoS) is maintained for all users assuming fixed and identical UL transmit power over all channels. In \cite{Hjoint}, a joint DL and UL aware cell association in HetNets is proposed, where each BS has a finite number of resource blocks to distribute among users. In \cite{Yopt}, BS association and channel assignment are jointly considered in both DL and UL in interference-limited HetNets, where biased BS association is based on highest SINR in DL, and when bias is identical for all BSs, the nearest BS is associated to the user. 

In \cite{Singh2}, SIR and uplink rate distribution are used for load balancing and power control in HetNets, where cell association is based on path-loss. In \cite{Vuplink}, the outage probability in UL in a two-tier CDMA-based network with shared spectrum is discussed, and a lower bound for the outage in femtocells is derived. Using stochastic geometry in \cite{Hon}, a simple model for obtaining the outage and calculating the spectral efficiency in UL in non-heterogeneous networks and HetNets is proposed. In \cite{Hon}, channel assignment is fixed and channel inversion is used for power control. In \cite{Mingyi}, a game-theoretic and distributed joint channel assignment and power control in HetNets is presented, but inter-cell interference is ignored.

Energy efficiency in the HetNets' UL is considered in \cite{Zhang}, where fractional power control is applied by each user to maximize its SINR subject to its maximum transmit power constraint. In \cite{Ding}, energy efficiency for all users in a HetNet is maximized, where each user is associated with a BS that satisfies its SINR with minimum UL transmit power. Considering the above, in this paper we address joint UL transmit power control, channel assignment and cell association in two-tier HetNets consisting of conventional macro BSs and femto BSs to maximize the users' UL sum rate, where each user should have at least one channel. This problem is NP-hard, with no closed form solution. We divide the problem into two sub-problems: one for power control and one for sub-channel assignment and cell association; and iteratively alternate between the two until convergence. 

The rest of this paper is organized as follows. In Section \ref{System Model}, the system model is described. In Section \ref{Problem Formulation}, the problem is formulated and divided into two sub-problems, followed by a description of our proposed algorithm. Section \ref{Numerical Results} contains simulation results, and conclusions are in Section \ref{Conclusion}.

\section{System Model}\label{System Model}

Consider a HetNet with a total of $M$ BSs consisting of one macro BS and $M-1$ femto BSs, $N$ users uniformly distributed in the entire coverage area, and $K$ sub-channels in each BS. In this network, two schemes for frequency reuse is possible: full reuse, and fractional reuse. In the former, all sub-channels are used in all BSs; and in the latter, some sub-channels are exlusively used by femto BSs, and the rest can be used by either the macro BS or femto BSs. The maximum transmit power for user $i$ is $p_i^{\mathrm{max}}$, and $p_{i,j,k}$ is the transmit power of user $i$ to BS $j$ on sub-channel $k$.

The gain in sub-channel $k$ between user $i$ and BS $j$ is
\vspace{-0.05 in}
\begin{equation}
 h_{i,j,k}=g_{i,j,k} f_{i,j,k},\;\, 1\leq i\leq N, \; 1\leq j\leq M, \; 1\leq k\leq K
\end{equation}
where $g_{i,j,k}=L(d_{i,j})+S$ is the impact of large scale slow fading for each sub-channel between user $i$ and BS $j$, and $f_{i,j,k}$ is the impact of small scale fast fading. Path loss between user $i$ and its BS $j$ is $L(d_{i,j}) = L^{\mathrm{fixed}}+ 10\alpha \log(d_{i,j})$, where $L^{\mathrm{fixed}}$ is the fixed path-loss and $d_{i,j}$ is the distance between user $i$ and BS $j$. Shadowing is denoted by $S$ and is a lognormal random variable with zero mean and $\sigma_\text{S}$ standard deviation. The small scale fast fading $f_{i,j,k}$ is a Rayleigh distributed random variable with zero mean and unit variance for all $i,j,k$. Considering the above, sub-channel gain $h_{i,j,k}$ is an exponentially distributed random variable \cite{Hjoint} with parameter $\lambda_{i,j,k}$, where
\vspace{-0.08 in}
\begin{equation}
\lambda_{i,j,k}= \frac{1}{E[h_{i,j,k}]}=\frac{1}{g_{i,j,k}}.
\end{equation}

The SINR in BS $j$ from user $i$ on sub-channel $k$ is
\vspace{-0.07 in}
\begin{equation}\label{si1}
\gamma_{i,j,k}=\frac{p_{i,j,k} h_{i,j,k}}{\sum_{l=1\setminus{i}}^N\sum_{s=1\setminus{j}}^Mp_{l,s,k}h_{l,j,k}+\Delta fN_0},
\end{equation}
where $N_0$ is the noise power, $\Delta f$ is the sub-channel bandwidth, $1\setminus{i}$ means all users except user $i$, $1\setminus{j}$ means all BSs except BS $j$, and $\sum_{l=1\setminus{i}}^N\sum_{s=1\setminus{j}}^Mp_{l,s,k}h_{l,j,k}$ is the interference on sub-channel $k$ in BS $j$ from all users in other BSs transmitting on the same sub-channel $k$. As in \cite{3gpp}, we also assume users can associate with more than one BS.
\section{Problem Formulation}\label{Problem Formulation}

We consider the users' sum rate as the utility function, and obtain the binary matrix {\bf{X}} and transmit power $p_{i,j,k}$ for utility maximization. The entries $x_{i,j,k}$ of matrix {\bf{X}} denote association of user $i$ on sub-channel $k$ to BS $j$. The optimization problem is  
\vspace{-0.07 in}
\begin{equation}\label{opt1}
\begin{array}{rlllll}
\displaystyle {\max_{\bf{X, P}} }& \multicolumn{1}{l}{\sum_{i= 1}^{N}\sum_{j = 1}^{M}\sum_{k = 1}^{K}x_{i,j,k}\log(1 + \gamma_{i,j,k})}, \\
\end{array}
\end{equation}
\vspace{-0.2 in}
\begin{eqnarray}
\textrm{Subject to} \left\{\begin{array}{c}
\hspace{-0.2 in}\textrm{C1:}\,\sum_{j=1}^{M}\sum_{k=1}^{K}x_{i,j,k}p_{i,j,k}\leq p_{i}^\text{max}, \qquad \forall i
\vspace{0.05 in}
\\
\hspace{-0.04 in}\textrm{C2:}\,\sum_{i = 1}^{N}x_{i,j,k}\leq 1,\qquad\qquad\qquad\qquad\, \forall j,k
\vspace{0.05 in}
\\
\hspace{-1.54 in}\textrm{C3:}\;x_{i,j,k}\in \{0 , 1\},
\end{array} \nonumber\right.
\end{eqnarray}
where $\gamma_{i,j,k}$ is as in (\ref{si1}). Similar to the transmit power matrix {\bf P}, the matrix {\bf{X}} is of dimensions $N\times M\times K$. The constraint C1 indicates that the sum of transmit power levels by user $i$ on all its sub-channels cannot exceed $p_i^{\mathrm{max}}$. The constraint C2 implies that each sub-channel in BS $j$ is used by only one user; and the constraint C3 indicates that assignment of sub-channel $k$ to user $i$ in BS $j$ is binary. Users can be associated with more than one BS at any given time.

This is a mixed integer and combinatorial NP-hard problem. To remedy its complexity, we divide it into two sub-problems: one for transmit power allocation to each user on different sub-channels, and one for sub-channnel assignment and BS association to each user. The two sub-problems are solved in sequence in each iteration until final convergence.

\subsection{Transmit Power Allocation}
We begin by setting all entries in the binary association matrix {\bf{X}} to 1 and proceed to obtain the transmit power $p_{i,j,k}$ for all values of $i,j,k$. The power allocation problem is
\begin{equation}\label{opt2}
\begin{array}{rlllll}
\displaystyle {\max_{\bf{P}} }& \multicolumn{1}{l}{\sum_{i= 1}^{N}\sum_{j = 1}^{M}\sum_{k = 1}^{K}x_{i,j,k}\log(1 + \gamma_{i,j,k})}, \\
\end{array}
\end{equation}
\hspace{0.5 in} Subject to  C1. \\
This problem obtains each user's transmit power on its sub-channels to maximize the network's sum-rate for all $x_{i,j,k}$. The water-filling algorithm solves this problem by increasing the transmit power on those sub-channels that experience less interference (see Appendix I), which maximizes the sum rate in the network. However, when multiple users are on one sub-channel, interference is caused. To mitigate this, after convergence of transmit power in each iteration, we apply the below scheme for sub-channel assignment and repeat this cycle until only one sub-channel is assigned to each user.

\subsection{Sub-Channel Assignment}
To satisfy C2, after transmit power convergence in each iteration, assignment of sub-channels in {\bf{X}} is pruned by removing a user from subchannels in each iteration until finally one user remains on each sub-channel. To maximize the sum rate, the below criteria is checked for all users in each iteration:
	\begin{equation}
	\delta(k')= \sum_{i= 1}^{N}\sum_{j = 1}^{M}\sum_{k = 1}^{K}{r_2}_{i,j,k}(k')-\sum_{i= 1}^{N}\sum_{j = 1}^{M}\sum_{k = 1}^{K}{r_1}_{i,j,k}(k'),
	\end{equation}
where ${r_1}_{i,j,k}(k')$ and ${r_2}_{i,j,k}(k')$ are the rates of user $i$ on sub-channel $k$ in base station $j$ before and after removing a user from sub-channel $k'$, respectively, and $\delta (k')$ is the increase in the users' sum rate caused by removal of a user from sub-channel $k'$. The user with the highest $\delta (k')$ is removed from sub-channel $k'$ unless only one sub-channel is assigned to that user. In this case, the user with the second highest $\delta (k')$ is removed. This implies that each user can transmit on at least one sub-channel.  

\begin{algorithm}
\caption{Joint Uplink Resource Allocation in HetNets}\label{A1}
\begin{algorithmic}[1]
\setstretch{0.85}
\setlength{\intextsep}{0pt}
\State Initialize: Set ${\bf{X}}={\bf{1}}$ and $p_{i,j,k}=p^{\mathrm{max}}/ (M\times K)$.
\Repeat
\parState{for each base station $j=1$ to $M$ }
\Repeat
\parState {for each sub-channel $k'=1$ to $K$}
\While { more than one user is on sub-channel $k'$}
\parState  {\hspace{-0.3cm}Calculate $\delta (k')$ in (6) for all users on sub-channel} 
\NoNumber{\hspace{-0.3cm}{$k'$, and choose the user with the highest $\delta (k')$}}
\If {that user has another sub-channel}
\parState{remove it from sub-channel $k'$}
\Else
\parState{remove the user with the next highest $\delta(k')$  who has more than one sub-channel from sub-channel $k'$}
\EndIf
\Repeat
\parState{\hspace{-0.3cm}update the transmit power for all users via (5)}
\Until{{\bf{P}} converges}
\EndWhile
\Until{$k'=K$}
\Until {$j=M$}
\end{algorithmic}
\end{algorithm}

\section{Numerical Results and Discussion}\label{Numerical Results}

A two-tier HetNet consisting of 1 macro BS, 4 femto BSs, and 25 users uniformly distributed in a 1000 m $\times$ 1000 m cell is considered. Without loss of generality the macro BS is placed at (0,0). There are 20 sub-channels in each BS, but the macro BS only uses some (not all) sub-channels in the fractional frequency reuse scheme. Since LTE sub-channels are 180 kHz wide, thermal noise power is set to -111.45 dBm. The maximum transmit power of each user is set to 20 dBm. The large scale path loss between each user and the macro BS is $L(d)$ = 34 + 40 log($d$), where $d$ is the distance in meters. The large scale path loss between each user and a femto BS is $L(d)$ = 37 + 30 log$(d)$. Besides, shadowing is modeled by a lognormal random variable with 8 dB standard deviation, and Rayleigh small scale fading is modeled by an exponentially distributed random variable with zero mean and unit variance as in \cite{Hjoint}. Simulation parameter values are as in Table I.

\begin{figure}[h!]
	\centering
	\includegraphics[scale = 0.5464]{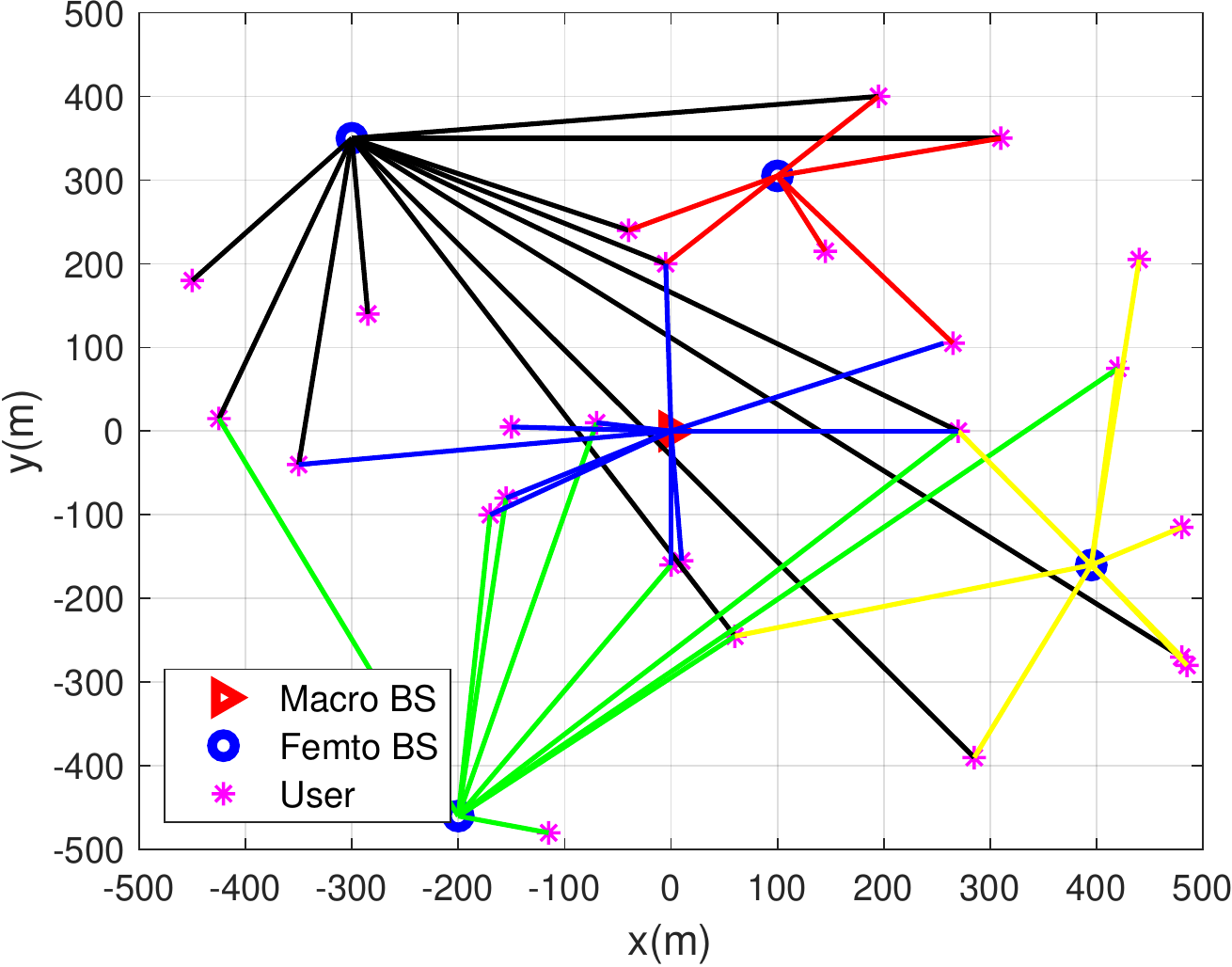}
	\caption{Initial user association.}
	\label{fig1}
\end{figure}
\begin{figure}[h!]
	\centering
	\includegraphics[scale = 0.5464]{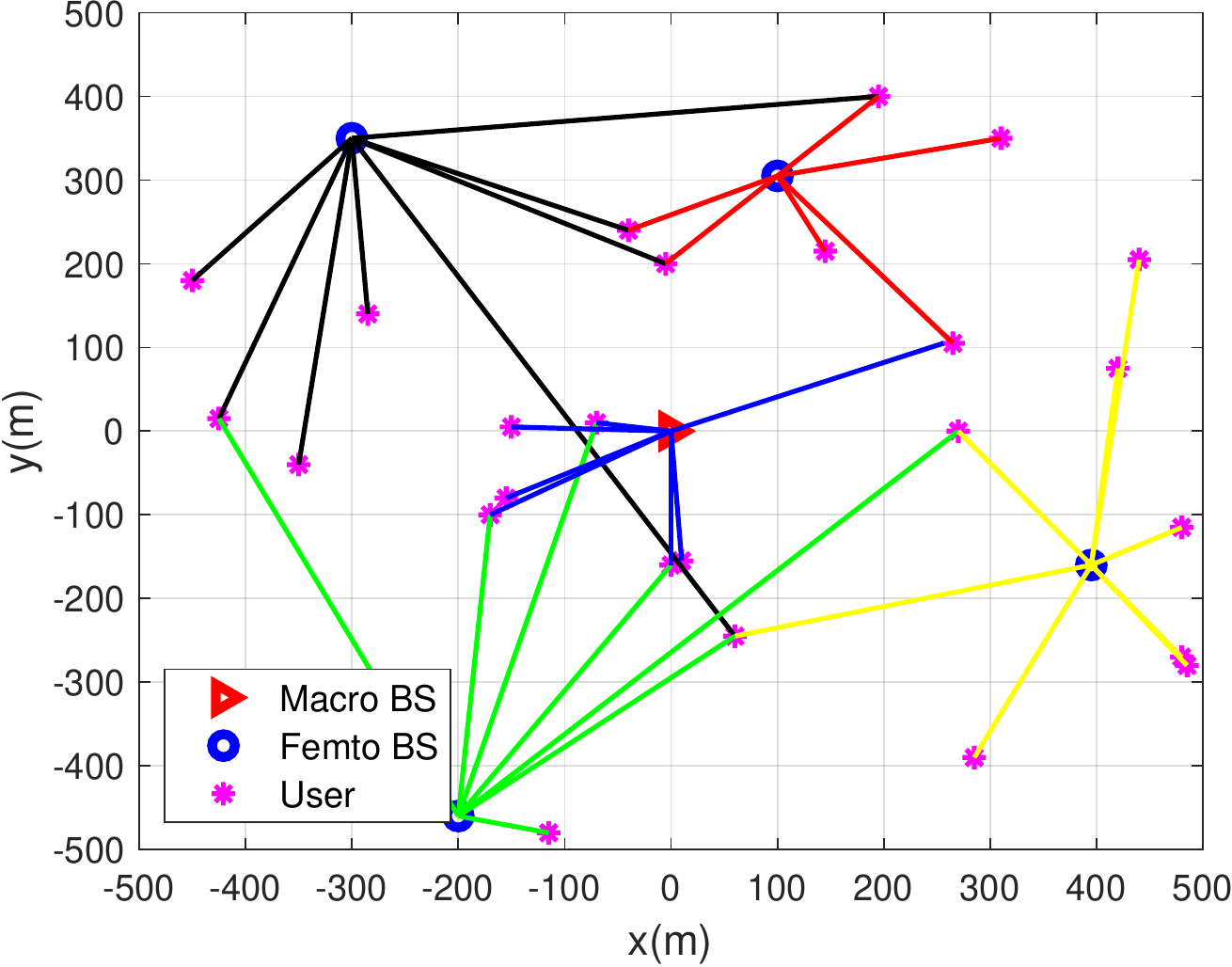}
	\caption{Final user association.}
	\label{fig2}
\end{figure}
\begin{figure}[h!]
	\centering
	\includegraphics[scale = 0.5464]{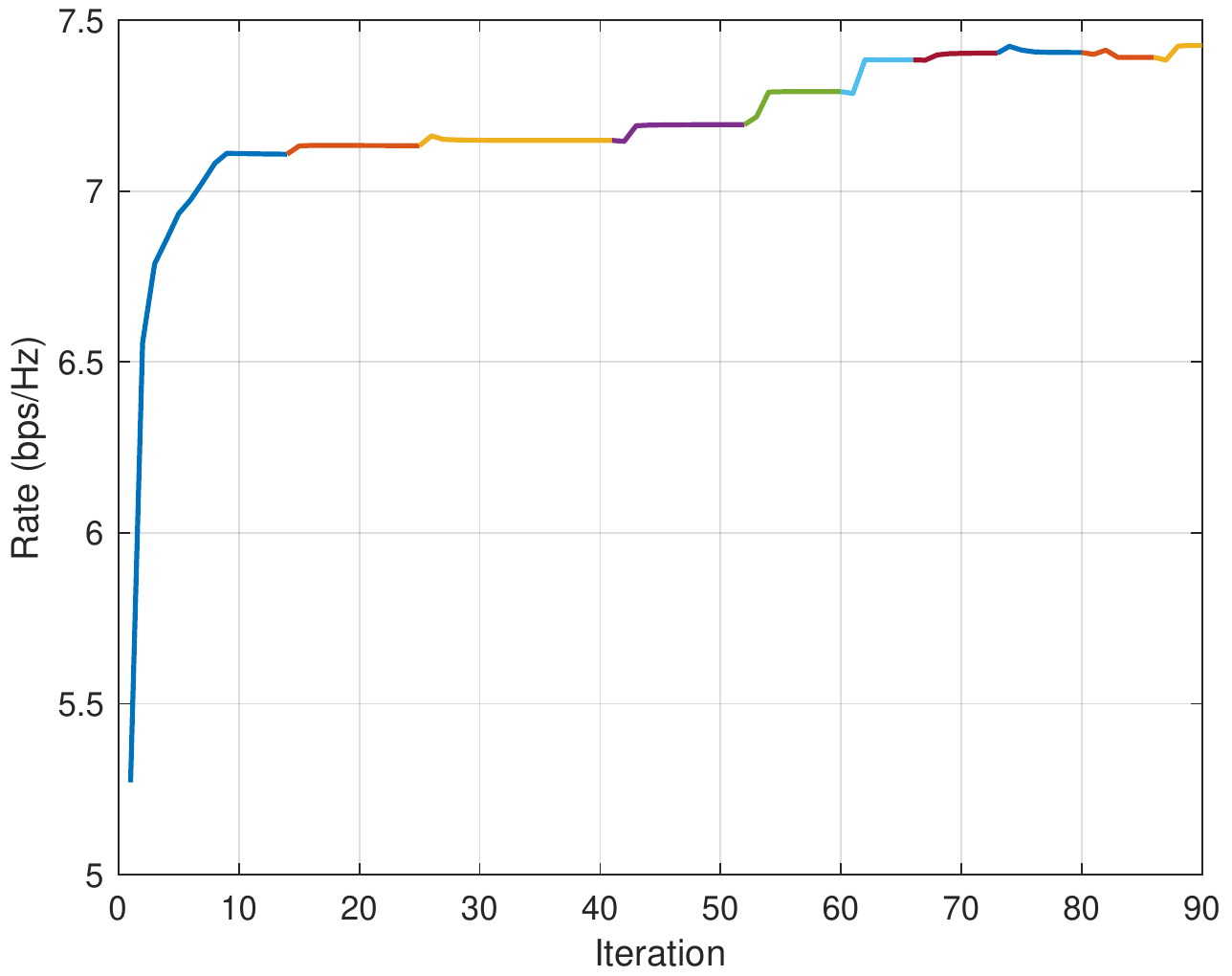} 
	\caption{Convergence of our proposed algorithm.}
	\label{fig3}
\end{figure}

Figs. \ref{fig1} and \ref{fig2} show the initial and final association of users to BSs, respectively. Note that the final association is more balanced, and that some users connect to more than one BS. Fig. \ref{fig3} shows the convergence of our proposed algorithm. Each horizontal segment shows transmit power convergence in each iteration. When a user is removed from a sub-channel, the power is updated until convergence. The algorithm ends when there is no sub-channel with more than one user.

\begin{figure}[h!]
	\includegraphics[scale = 0.5464]{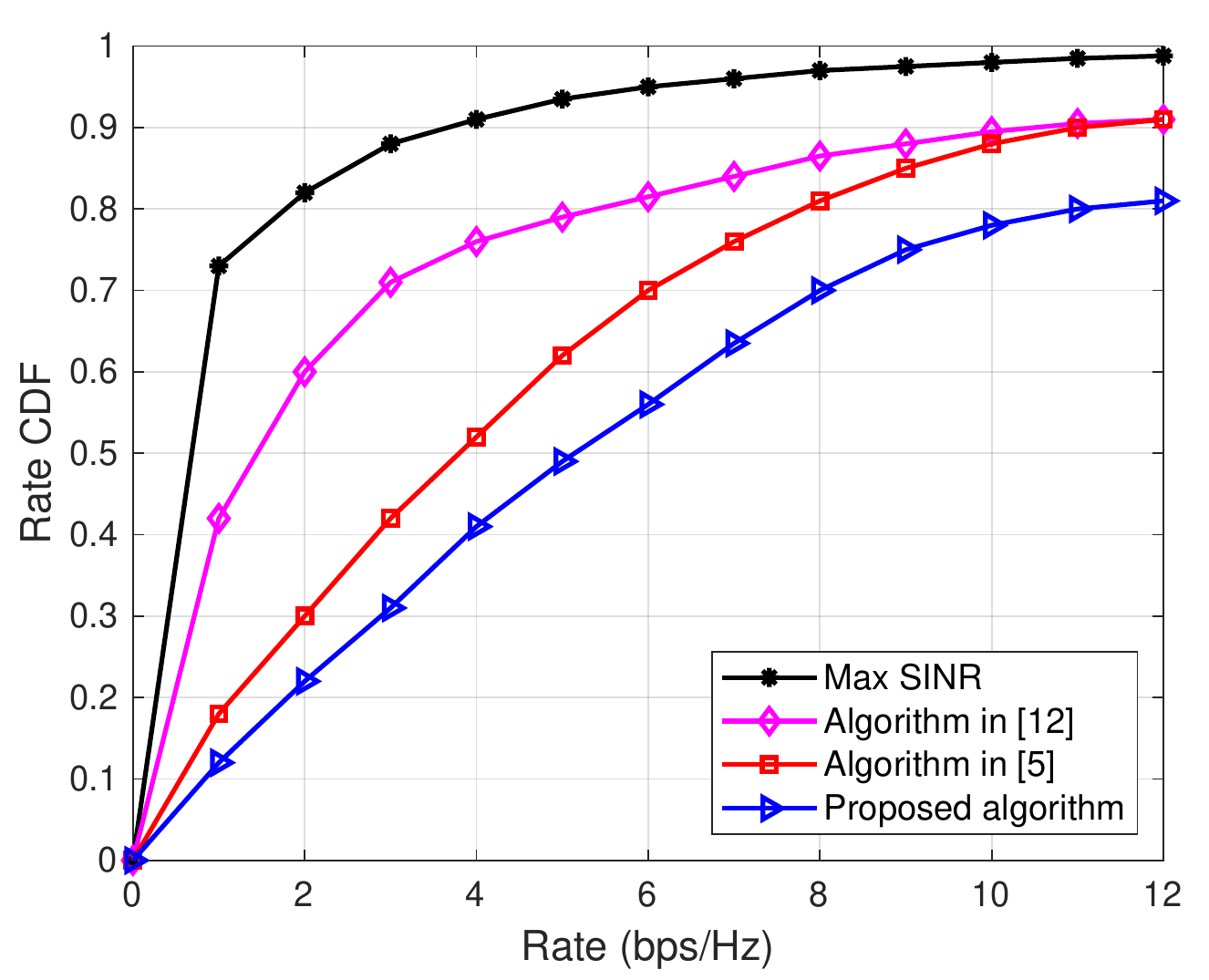}
	\centering
	\caption{Rate CDF for the max-SINR association scheme, the algorithm in  \cite{Hjoint} and our proposed algorithm.}
	\label{fig4}
\end{figure}

Fig. \ref{fig4} shows the cumulative distribution function (CDF) for the sum-rate in different schemes. 
In the max SINR association, sub-channels are equally distributed among users, and users transmit with equal power. Note that our scheme yields a transmission rate above 6 bps/Hz for almost 50\% of users as compared to 10\% of users in SINR-based schemes. Since our scheme applies transmit power control, it outperforms the algorithm in \cite{Hjoint} in which power control is not applied. Also, as can be seen in the same figure, the outage probability for $\gamma_\text{th}=0.6$ bps/Hz for the max SINR algorithm, for the method in \cite{Ding}, for the method in \cite{Hjoint}, and for our scheme is 42\%, 25\%, 10\%, and 7\%, respectively.

\begin{figure}[h!]
	\centering
	\includegraphics[scale = 0.5464]{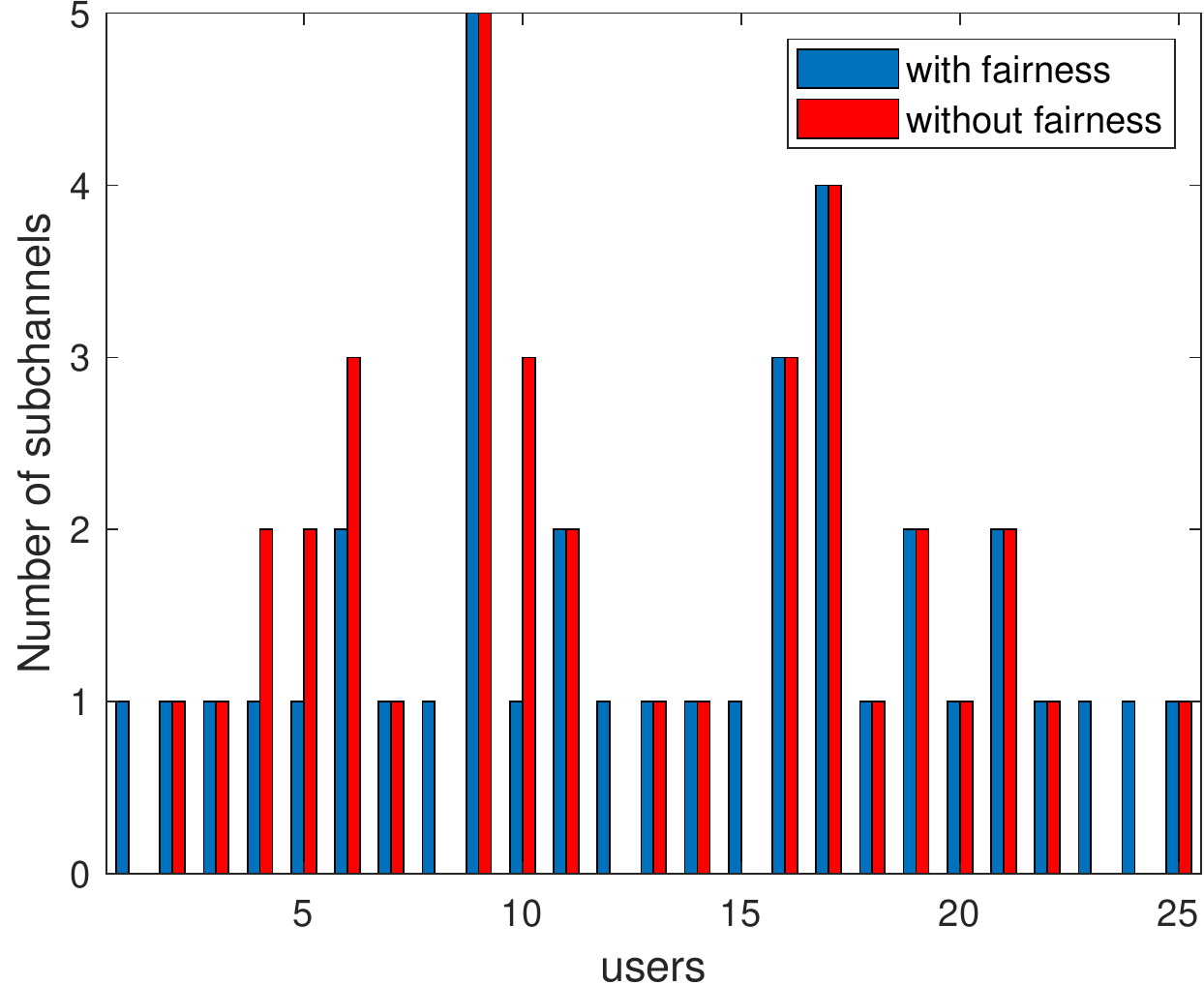}
	\caption{Sub-channel assignment with and without fairness.}
	\label{fig5}
\end{figure}

\begin{figure}[h!]
	\centering
	\begin{subfigure}{0.395\textwidth} 
		\includegraphics[width=\textwidth]{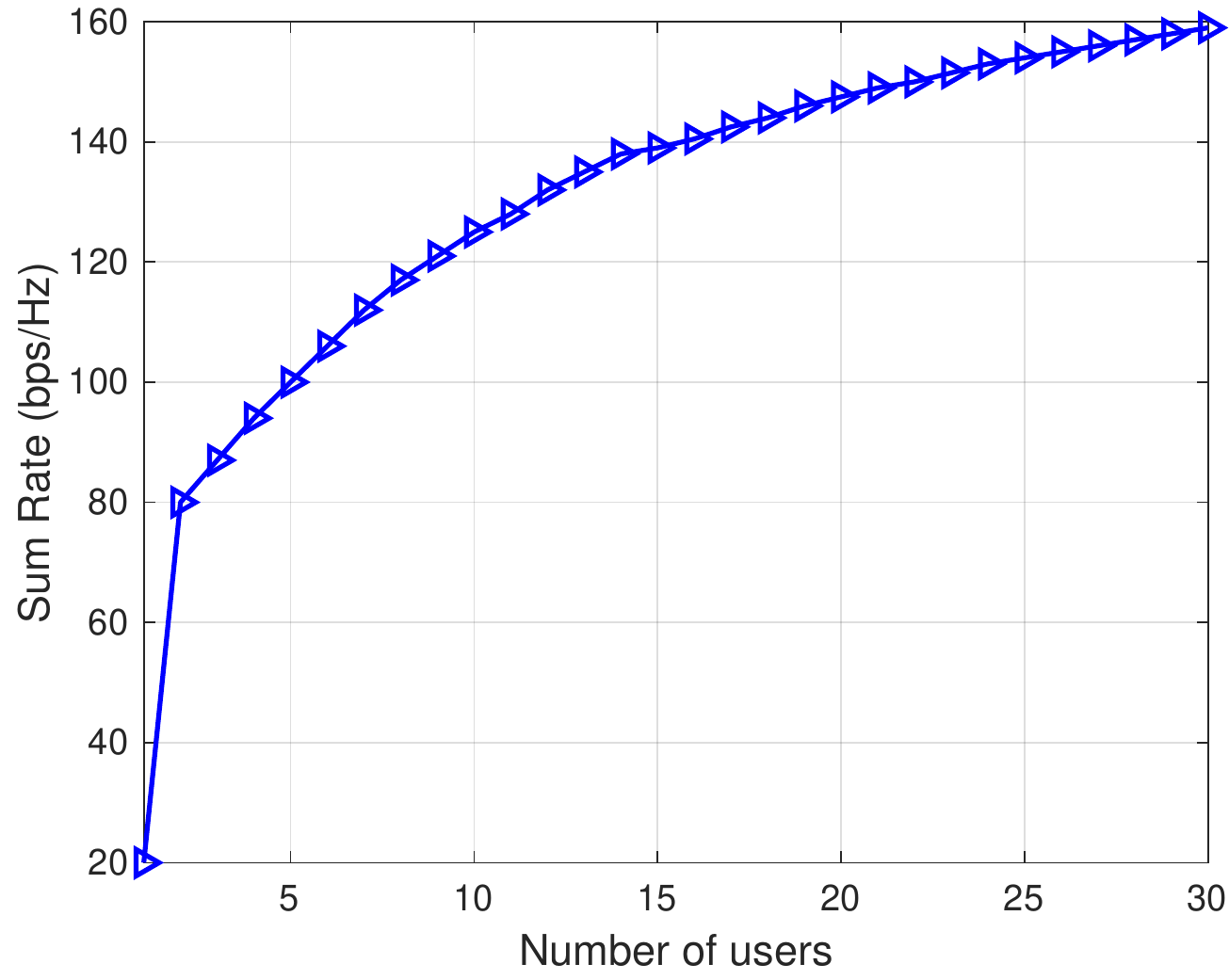}
		\caption{} 
		\label{6a}
	\end{subfigure}
	\vspace{1em} 
	\begin{subfigure}{0.395\textwidth} 
		\includegraphics[width=\textwidth]{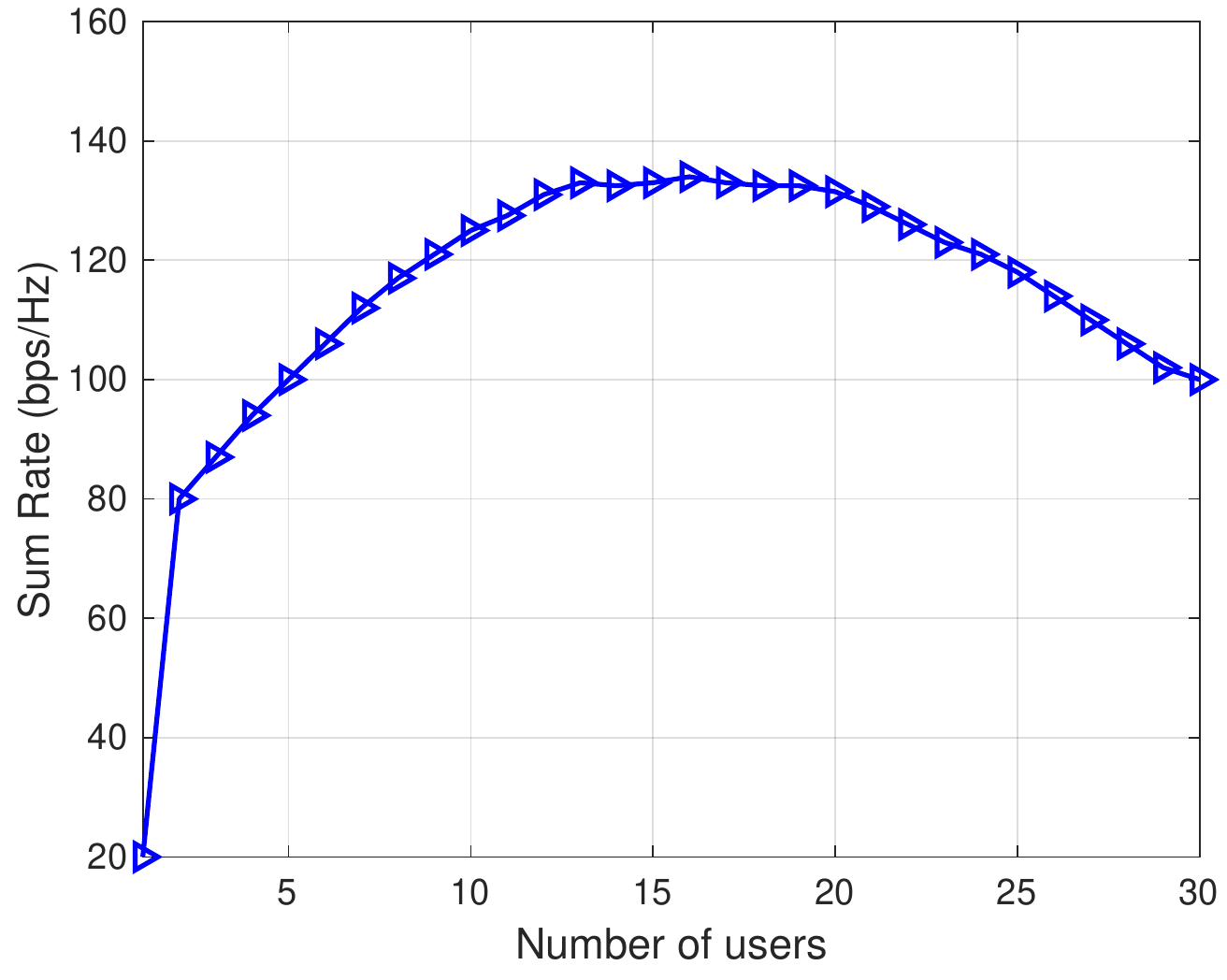}
		\caption{} 
		\label{6b}
	\end{subfigure}
	\caption{Sum rate: (a) some users do not have at least one sub-channel (fairness), (b) each user has at least one sub-channel.} 
	\label{fig6}
\end{figure}

Fig. \ref{fig5} shows the number of sub-channels assigned to each user with and without applying fairness (i.e., at least one sub-channe for each user), respectively. By increasing the number of users but without providing all users with at least one sub-channel, the sum rate is increased as in Fig. \ref{fig6}(\subref{6a}). This is because the probability of users with better channel increases. However, when all users are assigned at least one sub-channel, as can be seen in Fig. \ref{fig6}(\subref{6b}), increasing the number of users increases the sum-rate up to a point, after which the sum-rate is reduced. This is because fairness implies that in some cases a sub-channel is assigned to a user with a smaller gain. 

\begin{table}[t!]\label{j1}
	\caption{Simulation Parameters} 
	\vspace{-0.1 in}
	\centering          
	\resizebox{0.81\columnwidth}{!}{%
	\begin{tabular}{|l | c | c|}
		\hline     
		Parameter & Description & Value\\
		\hline \hline
		$N_0$ & Noise power & -111.45 dBm\\
		\hline
		$\sigma^2_{s}$&Shadow fading variance & 8 dB\\
		\hline
		$M$& Number of BSs & 5\\
		\hline
		$N$& Number of users & 25\\
		\hline
		$K$& Number of sub-channels & 20\\
		\hline
		$p_i^{\mathrm{max}}$& Fixed power consumption in each BS & 20 dBm\\
		\hline
		$\Delta f$&Sub-channel bandwidth & 180 KHz\\
		\hline
		$L^\mathrm{fixed}$&Fixed path loss for macro/femto users& 34/37 dB\\
		\hline
		$\alpha$& Path loss exponent for macro/femto users & 4/3\\
		\hline
	\end{tabular}%
}
\end{table}

\begin{figure}[h!]
	\centering
	\begin{subfigure}{0.395\textwidth} 
		\includegraphics[width=\textwidth]{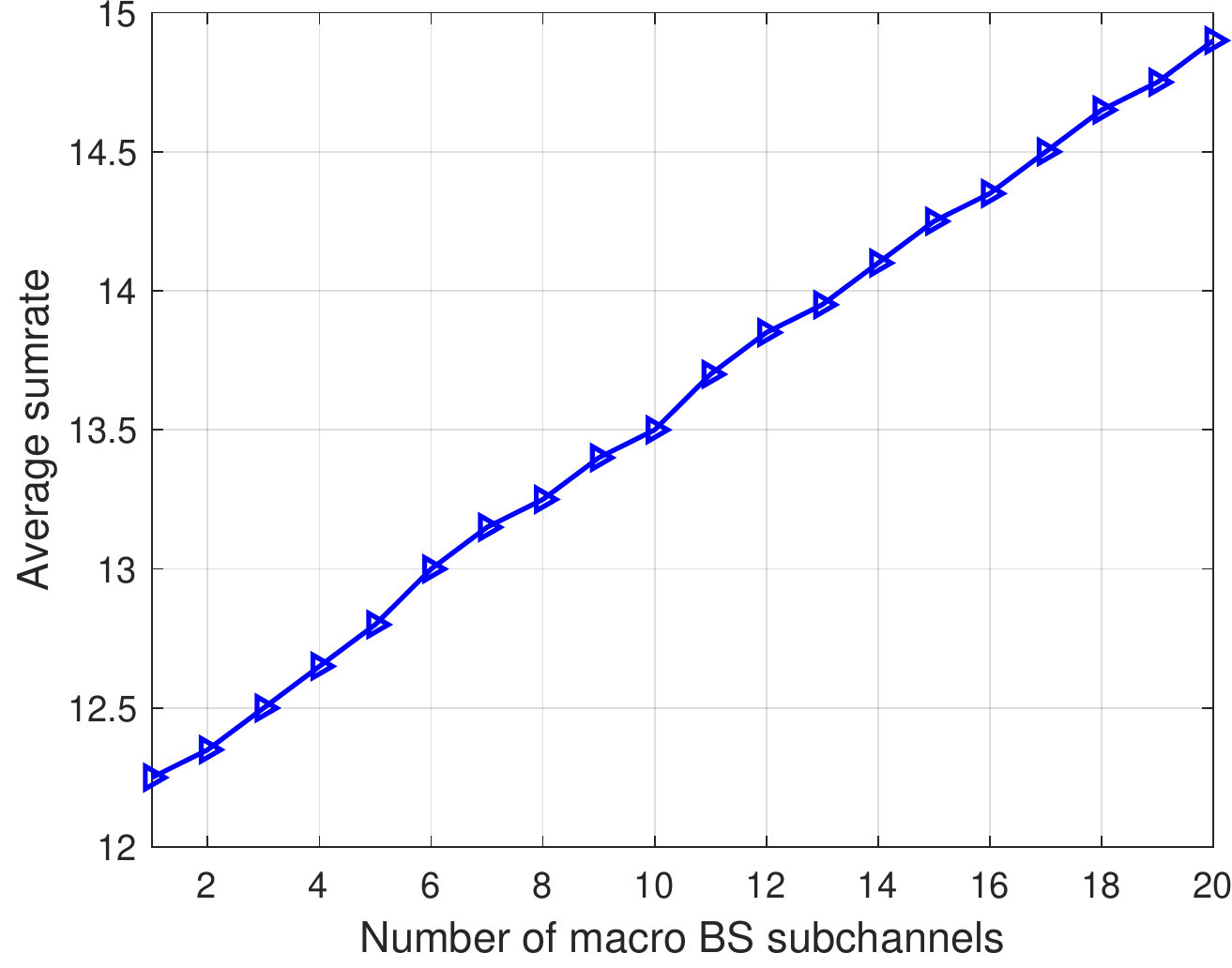}
		\caption{} 
		\label{7a}
	\end{subfigure}
	\vspace{1em} 
	\begin{subfigure}{0.395\textwidth} 
		\includegraphics[width=\textwidth]{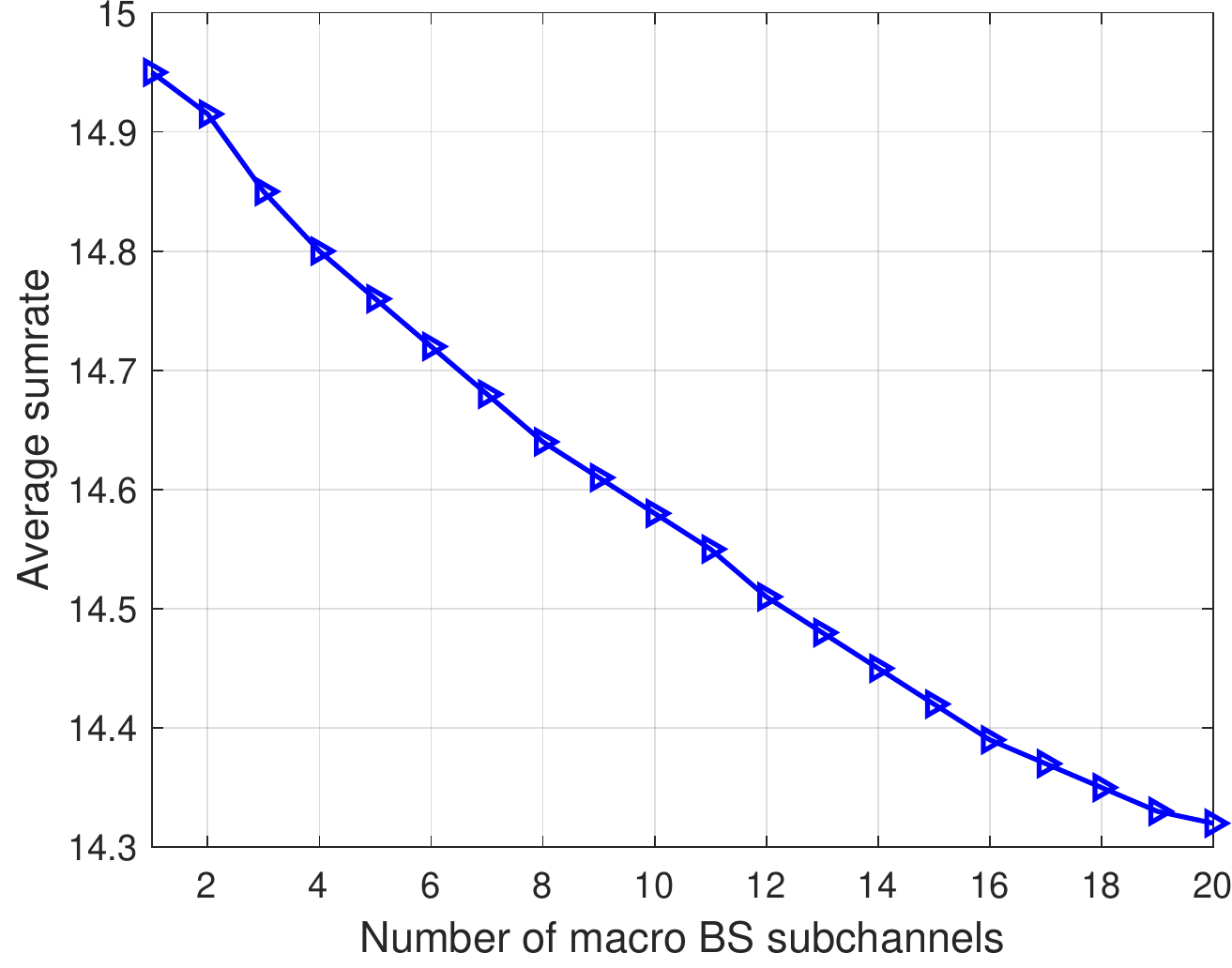}
		\caption{} 
		\label{7b}
	\end{subfigure}
	\caption{Average sum rate vs the number of macro BS sub-channels: (a) users are near BS, (b) users are far from BS} 
	\label{fig7}
\end{figure}

To study the impact of frequency reuse on the sum rate, we consider two cases. When many users are near the macro BS, making more sub-channels available to the macro BS increases the sum-rate, as in Fig. \ref{fig7}(\subref{7a}). In this case, full frequency reuse is better due to low interference. When many users are near femto BSs (indoor users), making more sub-channels available to the macro BS reduces the sum rate as in Fig. \ref{fig7}(\subref{7b}) due to interference caused by macro BS users on femto BS users. In this case, fractional frequency reuse should be applied.

\section{Conclusion}\label{Conclusion}

We proposed an algorithm for joint power control, sub-channel assignment, and BS association in the UL of two-tier HetNets when users' transmit power is bounded. The formulated optimization problem for maximizing the users' sum-rate is NP-hard with no closed form solution. We proposed to obtain sub-optimal solutions via our iterative algorithm, whose sum-rate is higher than those of the max SINR association and the algorithms in \cite{Hjoint} and \cite{Ding}. We showed that when a degree of fairness is applied by providing each user with at least one sub-channel, increasing the number of users increases the sum-rate up to a point, after which the sum rate is reduced. We also showed that when users are located near the femto BS, fractional frequency reuse is better than full frequency reuse.

\appendices

\section{Same Level Power Allocation}

The power allocation problem is
\begin{equation}\label{opt3}
\begin{array}{rlllll}
\displaystyle {\max_{\bf{P}} }& \multicolumn{1}{l}{\sum_{i= 1}^{N}\sum_{j = 1}^{M}\sum_{k=1}^{K}x_{i,j,k}\log(1 + \gamma_{i,j,k})}, \\
\end{array}
\end{equation}
\hspace{0.5 in} Subject to  $\sum_{j=1}^{M}\sum_{k=1}^{K}p_{i,j,k}\leq p_{i}^\text{max}, \: \:\: \:\forall i$. \vspace{0.05 in} \\ 
Assume $n$ sub-channels are assigned to user $i$ where interference plus noise on any sub-channel $k$ of user $i$ is $\xi_k$. We wish to distribute the transmit power of user $i$, denoted by $p_{i}^\text{max}$, to its sub-channels. This problem is formulated as
\vspace{-0.05 in}
\begin{equation}\label{opt4}
\begin{array}{rlllll}
\displaystyle {\max_{\bf{p}\geq0} }& \multicolumn{1}{l}{\sum_{k=1}^{K}\log(1+\frac{p_k}{\xi_k})}, \\
\end{array}
\end{equation}
\vspace{-0.2 in}
\begin{eqnarray}
\textrm{Subject to} \left\{\begin{array}{c}
\!\sum_{k=1}^{K}p_{k} = p_i^\text{max} \qquad\!\!\!\!\! \forall i,\\
p_k \geq 0\qquad\qquad\quad\, \forall k,
\end{array} \nonumber\right.
\end{eqnarray}
\vspace{-0.05 in}
where $\bf{p}$ is the transmit power vector whose elements are $p_k$.

\vspace{-0.05 in}
Instead of solving (\ref{opt4}), we solve its Lagrangian dual problem with $\lambda^*$ for the non equality constraint and $v^*$ for the equality constraint. We obtain the KKT conditions
\vspace{-0.05 in}
\begin{subequations}
\begin{equation}\label{9a}
\textbf{p}^* \geq \textbf{0}, \qquad
{\bf 1}^\text{T}\textbf{p}^*=p^{\mathrm{max}},
\end{equation}
\begin{equation}\label{9b}
\lambda^*\geq 0, \qquad
\lambda^*_kp_k^* = 0, \quad k=1,\cdots,K,
\end{equation}
\begin{equation}\label{9c}
-1/(1+\frac{p_k^*}{\xi_k})-\lambda_k^*+v^*=0, \quad k=1,\cdots,K.
\end{equation}
\end{subequations}
\vspace{-0.10 in}
Therefore,
\begin{subequations}
\begin{equation}\label{10a}
\textbf{p}^* \geq \textbf{0}, \qquad
{\bf 1}^\text{T}\textbf{p}^*=p^{\mathrm{max}},
\end{equation}
\begin{equation}\label{10b}
p_k^*[v^* - 1/(1+\frac{p_k^*}{\xi_k})] = 0, \qquad k=1,\cdots,K,
\end{equation}
\begin{equation}\label{10c}
v^* \geq 1/(1+\frac{p_k^*}{\xi_k}).
\end{equation}
\end{subequations}
When $v^*< 1$, (\ref{10c}) is satisfied only when $p_k^*> 0$. Hence, from (\ref{10b}), we write
\vspace{-0.1 in}
\begin{equation}\label{11}
v^* = 1/(1+\frac{p_k^*}{\xi_k}).
\end{equation}
Solving (\ref{11}) for $p_k^*$, we get $p_k^*= \xi_k/v^* - \xi_k$ when $v^*< 1$, and $p_k^*=0$ when $v^*> 1$, i.e., 
\vspace{-0.05 in}
\begin{equation}\label{133}
p_k^* = \mathrm{max}(0, \xi_k/v^* - \xi_k).
\end{equation}
By using (\ref{133}) in (\ref{9a}), we get
\vspace{-0.05 in}
\begin{equation}
\sum_{k=1}^{K}  \mathrm{max}(0, 1/v^* - \xi_k) = p^{\mathrm{max}}.
\end{equation}
The value of $1/v^*$ is the power level in the water filling algorithm, and forces the sum of transmit power levels in all sub-channels for each user to be $p^{\mathrm{max}}$\cite{boyd}.

\end{document}